# Autonomous interpretation of atomistic scattering data

Andy S. Anker[1,2]*, John L. A. Gardner[1], Louise A. M. Rosset[1], Andrew L. Goodwin[1], Volker L. Deringer[1]

[1]Department of Chemistry, University of Oxford, Oxford OX1 3QR, United Kingdom

[2]Department of Energy Conversion and Storage, Technical University of Denmark, DK-2800 Kgs. Lyngby, Denmark

*Corresponding author. Email: andy.anker@chem.ox.ac.uk / ansoan@dtu.dk

**Abstract**

Materials with bespoke properties have long been identified by computational searches, and their experimental realisation is now coming within reach through autonomous laboratories. Scattering experiments are central to verifying the atomic structures of autonomously synthesised materials. Yet, interpreting these measurements typically requires user expertise and manual processing, or machine learning (ML) models trained on predefined datasets, limiting fully autonomous materials discovery. Here, we introduce a differentiable optimisation framework that treats scattering calculations, energetics, and chemical constraints as a unified refinement problem. Capability demonstrations across molecules, crystal structures, nanoparticles, and amorphous matter show that this data-driven approach resolves structural degeneracies with multi-modal inputs – suggesting its usefulness for informing, and ultimately guiding, the operation of autonomous laboratories.

## Introduction

For over a century, scattering techniques have been essential for structural characterisation across virtually all classes of materials.(*1-3*) Today, scattering is not only established as a universal characterisation method but also increasingly being automated within autonomous laboratories.(*4, 5*) Yet, despite these advances, there remains no general way to automatically extract atomic structures from scattering data. In parallel, the continued advancement of widely-applicable ML-based interatomic potentials holds promise for the computational discovery of stable materials with tailored properties.(*6, 7*) These dual advances, in automated scattering experiments and in computational modelling, motivate efforts to realise computational predictions using autonomous laboratories.(*5*) The long-term vision is that researchers can specify desired properties, discover candidate materials computationally, and then produce them experimentally through autonomous synthesis.

A cornerstone of such workflows is scattering experiments to verify the atomic structures of synthesised materials. Methods such as small-angle scattering (SAS), powder diffraction (PD), or total scattering (TS) with pair distribution function (PDF) analysis can characterise materials across the full structural spectrum from well-ordered crystals to nanocrystalline, disordered, and fully amorphous phases.(*1-3*) Accordingly, protocols for automated interpretation of scattering data must operate robustly across this entire range. Each structure gives rise to a distinct pattern (*the forward calculation*); however, the reverse task of inferring an unknown structure from scattering data (*the inverse problem*, Figure 1A) is fundamentally ill-posed and far more challenging.(*8*) A fundamental complication arises because multiple distinct atomic arrangements can produce indistinguishable scattering patterns—known as the *uniqueness problem* (Figure 1A).(*9, 10*) Moreover, even high-quality scattering data will contain experimental noise, further complicating efforts to uniquely identify structures, and often requiring complementary data sources or constraints.

Several methods have emerged to address the inverse scattering problem, each with distinct strengths and limitations (see Supplementary Information (SI) section A, for a conceptual hierarchy). One approach refines user-defined structural models against scattering data, allowing extensive chemical knowledge integration but lacking a fully data-driven nature.(*11, 12*) This approach can be extended by combining multiple experimental and computational modalities but remains dependent on predefined structural assumptions.(*13*) Multi-modal extensions of molecular dynamics (MD) have been explored, where energies and forces are augmented with experimental data to guide the dynamics.(*14, 15*) Data-driven approaches such as reverse Monte Carlo (RMC) modelling iteratively adjust atomic positions to match experimental scattering data,(*16, 17*) but may require structural constraints for physically meaningful outcomes. Empirical potential structure refinement(*18*) and hybrid RMC(*19*) similarly refine structures against scattering data while incorporating interatomic potentials, improving physical plausibility but requiring effort from the user.

Recently, ML approaches have emerged to analyse scattering data by directly mapping experimental data onto sets of atomic coordinates.(*20, 21*) In these methods, models are first trained on large datasets of atomic structures paired with simulated scattering patterns. Although this step is computationally demanding, it is a one-time cost. Once trained, the model is fixed and can rapidly predict structures from new scattering data (Figure 1B). However, ML models remain fundamentally constrained by the span of their training sets, limiting their reliability for structures from uncharted chemical spaces. Consequently, automated interpretation of scattering data—particularly for materials outside the training distribution—remains unresolved in autonomous laboratories.(*21, 22*) An ideal methodology for autonomous scattering data interpretation would instead be (1) fully data-driven while being capable of: (2) handling structurally diverse systems; (3) identifying multiple structural solutions when uniqueness is lacking; and (4) accommodating multi-modal experimental inputs.

Here, we present a methodology specifically designed to meet these four requirements for the autonomous interpretation of multi-modal scattering data (Figure 1C). The central conceptual advance is the formulation of "data-to-structure" as a differentiable optimisation problem: scattering calculations, energetics, and chemical constraints are all embedded within a unified structural solution landscape, through which backpropagation drives refinement directly from experimental inputs. We demonstrate our framework across five illustrative case studies, ranging from molecules and nanoparticles to crystalline materials, and the canonical amorphous network of silicon.

**A unified differentiable optimisation framework for interpreting scattering data**

The starting point for our approach is to generate multiple candidate atomic structures ('walkers', here generated randomly). It then proceeds through an optimisation process exploiting modern deep learning optimisation tools. Here, each structure is refined on multi-modal scattering data (e.g., X-ray or neutron PD, SAS, TS, PDF), and multi-modal constraints (energy-, minimum interatomic distance-, atomic environment similarity-, and symmetry constraints) using gradient-based optimisation, here carried out via automated differentiation and the Adam optimiser(*23*). Simulated scattering patterns are computed for each candidate structure, and their agreement with experimental datasets, alongside energetic and chemical constraints, is quantified into a single, unified loss value. Structural updates occur through backpropagation using a combination of local and global optimisation schemes mitigating the risk of walkers becoming trapped in local minima (section B, SI). The methodology is implemented in a software package that we call Scatter2xyz (see Methods, SI for details).

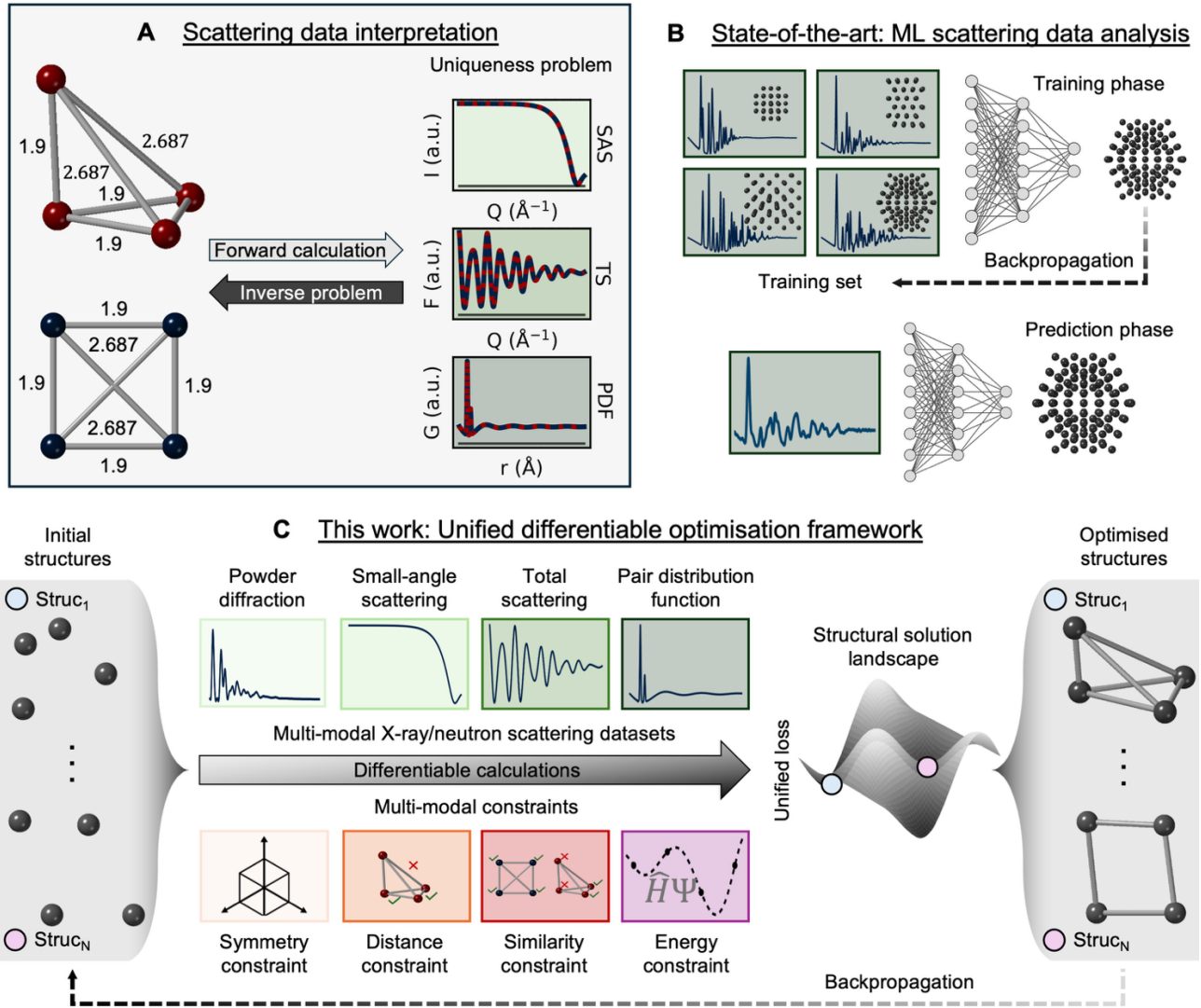

**Figure 1 | Interpreting scattering data through a unified differentiable optimisation framework.** (**A**) The scientific process of scattering data interpretation can be formulated as the *forward* (known structure → simulated scattering data) and *inverse* (experimental scattering data → structure) problems. While the forward calculation is straightforward, the inverse problem suffers from the *uniqueness problem*: distinct atomic arrangements can yield indistinguishable scattering patterns, as illustrated by triangular- and square-based four-atom configurations producing identical SAS, TS, and PDF signals, drawn following Ref. (*10*). (**B**) ML approaches address the *inverse problem* by training on large datasets of structure–scattering pairs (*training phase*), a computationally expensive but one-time cost. Once trained, the model is fixed and acts as a rapid predictor for new scattering patterns (*prediction phase*). This enables fast interpretation but constrains applicability to the span of the training set. (**C**) Our differentiable optimisation framework formulates scattering interpretation as a unified refinement problem. Multiple candidate structures are initialised and iteratively optimised against multi-modal scattering datasets (PD, SAS, TS, PDF) under chemical/energetic constraints (symmetry, distance, similarity, energy). Gradient-based backpropagation across the structural solution landscape drives optimisation, producing models consistent with experimental data while explicitly exposing degeneracies through multi-modal inputs.

To illustrate the methodology, we begin with simulated scattering data for a simple species of four P atoms (Figure 1A) and systematically examine how different input modalities affect the optimisation outcomes. In the first scenario (Figure 2A–I), randomly generated atomic arrangements are optimised solely with respect to energetic stability (minimum energy, evaluated with a ML interatomic potential trained on various DFT-labelled $P_{1-4}$ structures, see Methods, SI). The resulting structural solution landscape (interpolated from the best structures identified among 100 walkers) reveals a global minimum corresponding to a low-energy tetrahedral structure (Figure 2B, pink). Although energetically favourable, this arrangement does not describe the fictitious scattering patterns (Figure 2C). Conversely, when synthetic SAS, TS, and PDF data are provided along with minimal distance constraints (Figure 2–II), optimisation results in either a square (Figure 2B, blue) or a triangular configuration (Figure 2B, red). Although these configurations are energetically less favourable, both now perfectly reproduce the scattering pattern (Figure 2D). Notably, the walkers populate the structural solution landscape around these two distinct global minima, highlighting the inherent structural ambiguity—the '*uniqueness problem*'—when relying solely on scattering data. In an autonomous laboratory setting, such unresolved ambiguities can serve as a diagnostic signal, prompting the system to acquire additional experimental data streams. For example, incorporating an atomic environment similarity metrics (inspired by NMR) favours the square configuration (Figure 2A–III), while inclusion of energetic constraints directs the optimisation towards the triangular arrangement (Figure 2A–IV). Thus, a unified differentiable optimisation framework meets the outlined requirements for autonomous laboratories by being fully data-driven, handling structurally diverse systems, identifying multiple plausible solutions, and handling structural ambiguities by integrating multi-modal experimental inputs.

Beyond atomic positions, composition itself can vary in practical synthesis workflows. In an autonomous laboratory, it is therefore also essential that the optimisation framework can refine composition. Figures 2A–V illustrate this capability by moving beyond the assumption of a known stoichiometry. The framework begins from random atomic species and coordinates, yet still recovers either the square or triangular configuration along with the correct composition (P$_4$). Again, walkers span the structural solution landscape around these two global minima. In principle, the framework can also remove atoms, however, the more demanding the optimisation task (such as jointly refining number of atoms, composition, and positions), the greater the value of complementary data streams in guiding convergence (section C, SI). For example, elemental analysis could provide compositional constraints, making it easier for the optimisation to reach chemically realistic solutions while retaining a fully data-driven refinement process.

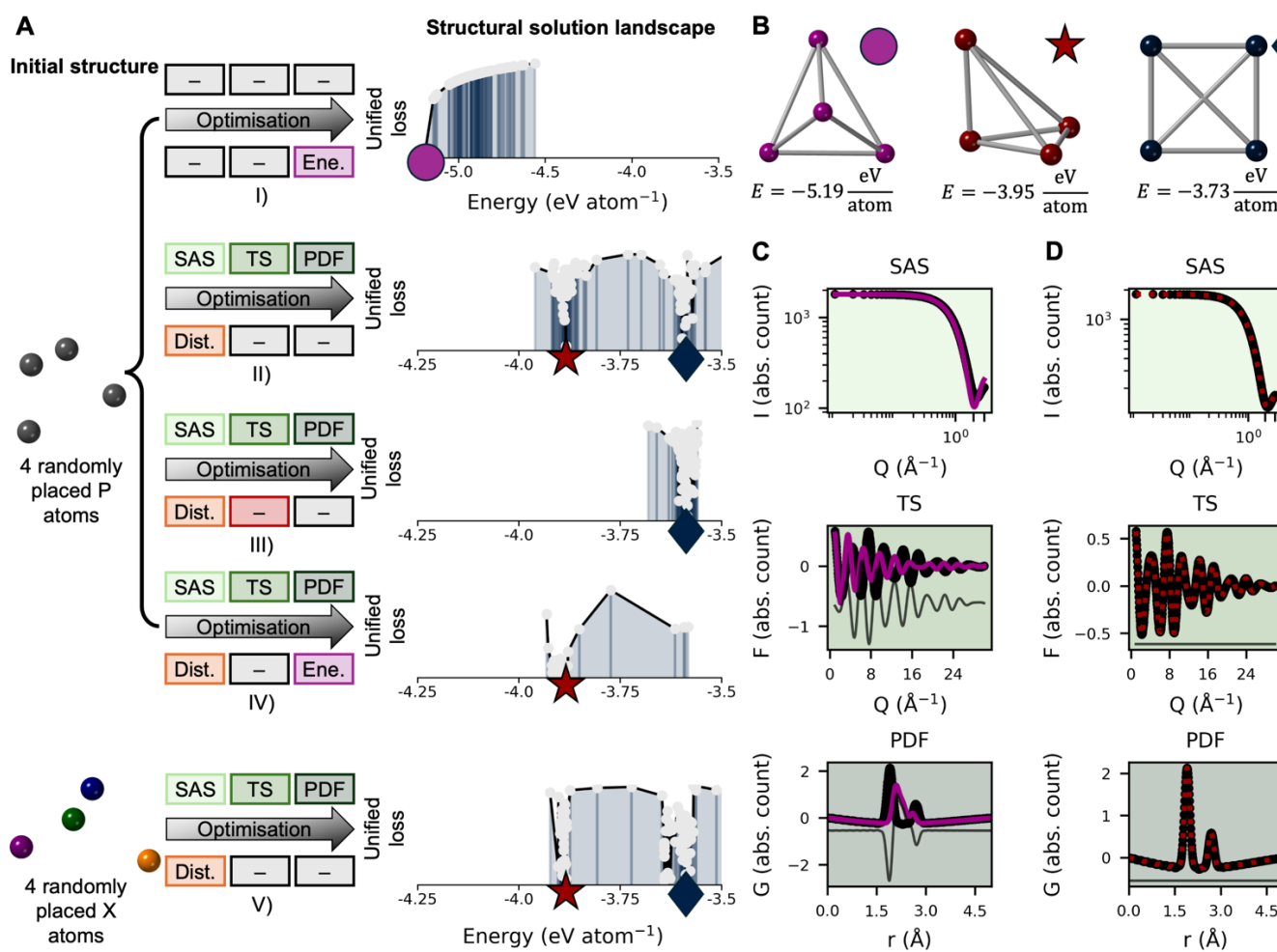

**Figure 2 | Illustrative examples of resolving structural ambiguities using multi-modal constraints. (A)** Five scenarios illustrating how random atomic configurations evolve into distinct structural solutions (the landscape is interpolated from the best solutions identified among 100 walkers) when optimised against different

combinations of input data and constraints. (**I–IV**) Four randomly placed P atoms optimised against I) lowest energy, (**II**) combined SAS, TS, and PDF data with distance constraints, (**III**) combined SAS, TS, and PDF data with distance and atomic similarity environment constraints, and (**IV**) combined SAS, TS, and PDF data with distance and energy constraints. (**V**) optimisation of both composition and coordinates; four randomly placed atoms of random species are refined against SAS, TS, and PDF data with distance constraints. (**B**) The principal structural motifs identified across scenarios, with their associated energies and scattering pattern fits. (**C–D**) Example I produces the lowest energy structure but fails to reproduce the scattering data, whereas examples II–V provide excellent agreement with experimental patterns.

**Case studies**

*Crystalline materials*

For crystalline materials, benchmarks of ML methods on the MP-20-PXRD dataset (*7, 24*) report $R^2_{wp}$ values of ~32% using diffusion models, whereas values below 10% are generally considered successful (see Methods, SI).(*25*) Figures 3A–B show that the present approach achieves $R^2_{wp}$ values below this threshold for crystalline silicon (c-Si, diamond structure) and $CeO_2$. In Figure 3A, our approach generates a structural model of c-Si from simulated PD data, achieving an $R^2_{wp}$ of 6%. Here Scatter2xyz finds the correct solution, albeit in a nonstandard crystallographic setting that is equivalent to the conventional $Fd\bar{3}m$ description. Figure 3B similarly shows Scatter2xyz generating a fluorite-type $CeO_2$ structure from either X-ray *($R^2_{wp}$ = 0%)* or neutron ($R^2_{wp}$ = 10%, section D, SI) PD data under a cubic symmetry constraint. These examples demonstrate that gradient-based optimisation within a differentiable framework can generate crystallographically meaningful structures without training-set bias. ML methods, by contrast, first require training on large sets of structure–scattering pairs, after which the trained model can act as a near-instant predictor (Figure 1B). For instance, in our own earlier work, a graph-based conditional variational autoencoder was trained on 3742 datasets in ~14.5 GPU-hours, after which predictions took less than a second.(*26*) Such speed is convenient, but applicability is fundamentally tied to the span of the training set; in this example for monometallic structures of up to 200 atoms. The differentiable optimisation framework instead formulates scattering interpretation as a general refinement problem (Figure 1C), agnostic to chemical space or structural libraries. While this generality comes with higher computational cost (minutes, section E, SI), it provides the structural universality required for autonomous laboratories that can explore varied chemistries.

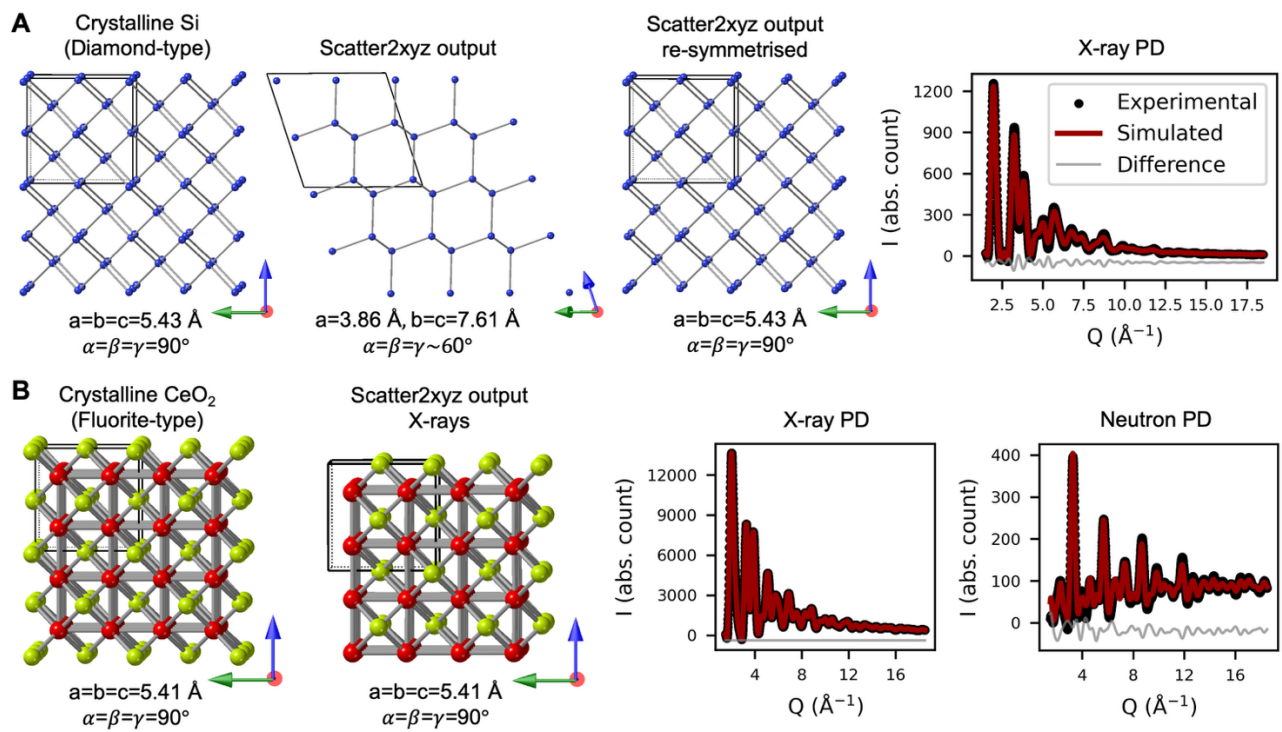

**Figure 3 | Crystalline structure generation from scattering data.** (**A**) Structural optimisation of the c-Si (diamond-type) structure. Starting from simulated X-ray PD data, the framework produces structural model that, while initially found in a nonstandard setting, can be re-symmetrised into the conventional $Fd\overline{3}m$ space group. The optimised model reproduces the scattering pattern with high fidelity. (**B**) Structural optimisation of the $CeO_2$ (fluorite-type) structure. Using simulated X-ray and neutron PD data under a cubic symmetry constraint, the framework converges on a fluorite structure that accurately matches the corresponding scattering signals.

## *The $C_{60}$ buckyball*

Within the nanostructure community, the canonical algorithmic challenge is the structure solution of buckyball $C_{60}$ from its TS or PDF pattern.(*27-29*) Figure 4–I demonstrates that the a configuration closely resembling that of the buckyball can be obtained in a fully data-driven manner, starting only from scattering data (SAS, TS, PDF) and a minimal distance constraint inferred directly from the PDF. The configuration is simultaneously consistent with the scattering patterns and energetically favourable. When excluding SAS data (Figure 4–II), the buckyball structure is not fully recovered within the same optimisation timeframe. Similarly, normalising scattering intensities (see Methods, SI)—though common in practice—further reduces available structural information by obscuring absolute intensity values (Figure 4–III). Both scenarios illustrate how reduced or incomplete data complicate autonomous structure analysis; while the generated structures provide information about shape, size, and local atomic environments, the structural models are not of the same quality as in Figure 4–I. In an autonomous laboratory context, such partial convergence may indicate either a complex structural solution landscape—necessitating longer optimisation times—or insufficient experimental information. Thus, limited convergence can indicate that additional data modalities are required, prompting further automated measurements.

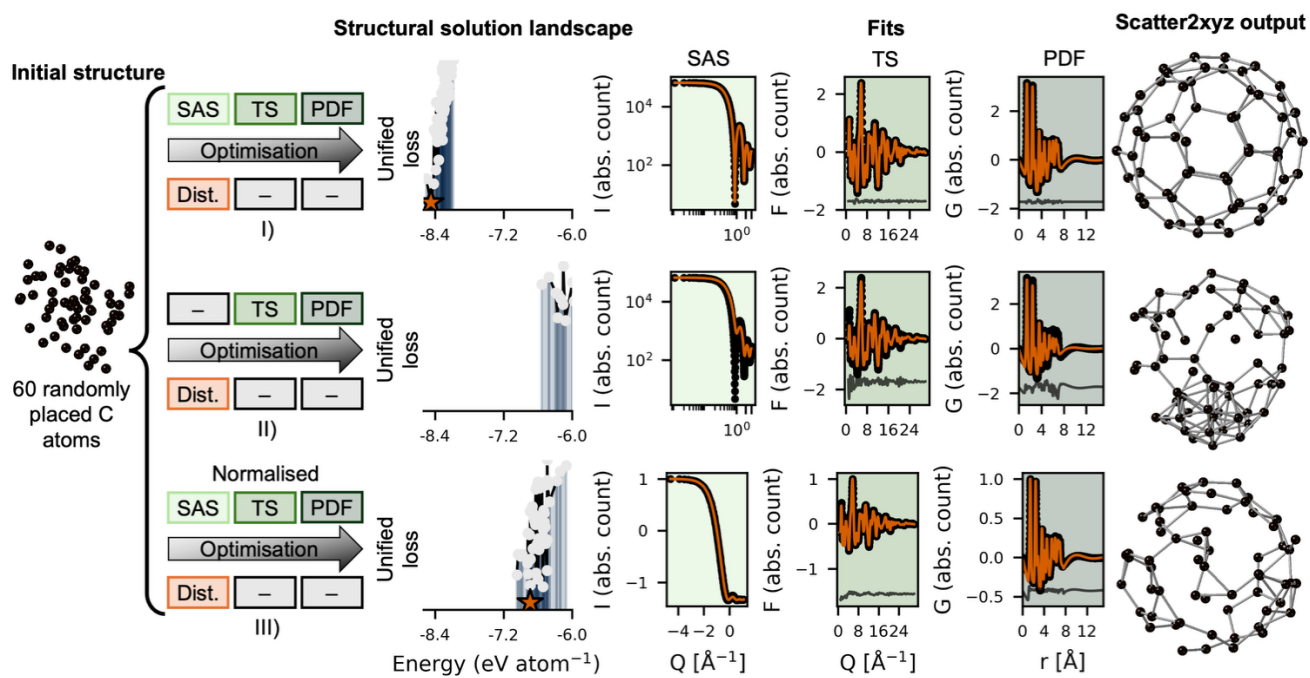

**Figure 4 | Interpretation of scattering data from the C₆₀ buckyball.** Examples illustrate how $C_{60}$ buckyball-like structures can be recovered starting from 60 randomly positioned carbon atoms. (**I**) using SAS, TS, and PDF data with distance constraints; (**II**) using only TS and PDF data with distance constraints; and (**III**) using normalised SAS, TS, and PDF data with distance constraints. The unified loss versus energy landscapes are interpolated from the best structural solutions among 50 walkers, with energies predicted using the MACE-MP0-a-large interatomic potential model.(*30*)

*Amorphous silicon*

a-Si is widely employed in solar-cell heterojunctions and thin-film transistors owing to its larger band gap compared to c-Si,(*31, 32*) and remains one of the most extensively studied disordered network solids.(*33-36*) Generally, a-Si is approximated as a continuous random network with minimal deviation from fourfold atomic coordination.

RMC modelling of a-Si can yield unphysical structures unless guided by appropriate constraints such as interatomic similarity criteria.(*16, 17*) MD simulations, employing quantum-mechanically accurate ML interatomic potentials,(*37-39*) have emerged as a state-of-the-art approach for generating accurate a-Si models that reproduce the experimental scattering pattern. In MD, structural models arise from physically motivated trajectories such as melt–quench protocols. In contrast, our unified differentiable framework does not simulate such processes but instead searches the structural solution landscape directly, optimising atomic structures against scattering data and energetics without pre-imposed protocols (Figure 5A). Due to GPU memory limits, the computations were performed for a relatively small simulation box (1000 atoms). Despite this limitation, the framework generates structures consistent with experimental scattering patterns and realistic energetics. We benchmark Scatter2xyz-derived models against a reference a-Si structure produced via melt–quench MD simulations, which yield energetically realistic structures with coordination numbers and bond-angle distributions characteristic of a-Si (Figure 5, blue).(*39, 40*) For clarity, we discuss energies as excess ($\Delta E$) relative to c-Si (diamond-type). Experimentally, a-Si exhibits a heat of crystallisation around $0.12 - 0.16$ eV atom$^{-1}$ for annealed and as-deposited samples, respectively.(*40, 41*) This experimental enthalpy is often approximated to the excess energy relative to diamond-type Si.(*42*) We use 0.2 eV atom$^{-1}$ as a target energy for the a-Si network. We conducted three Scatter2xyz runs, varying the weighting between energy and scattering data terms in the unified loss. Scatter2xyz produces structural models that describe experimental scattering patterns more accurately than MD-derived models (Figure 5A) while maintaining energies around

~0.2 eV atom$^{-1}$ above c-Si. However, direct structural interpretations reveal unphysical coordination numbers and bond-angle distributions (Figure 5B). A brief, 10-ps MD annealing simulation in the NpT ensemble at 500 K, substantially improves these local structural features, producing physically sensible bond angles and coordination numbers, though at the cost of a slightly compromised description of the scattering data (Figure 5C–D). This result exemplifies another manifestation of the *uniqueness problem*: multiple atomic arrangements (Figure 5E) can yield similar scattering patterns, but only some configurations represent physically realistic structures. Our framework addresses this by integrating interatomic potentials during optimisation, ensuring chemically realistic features such as coordination numbers and bond angles.

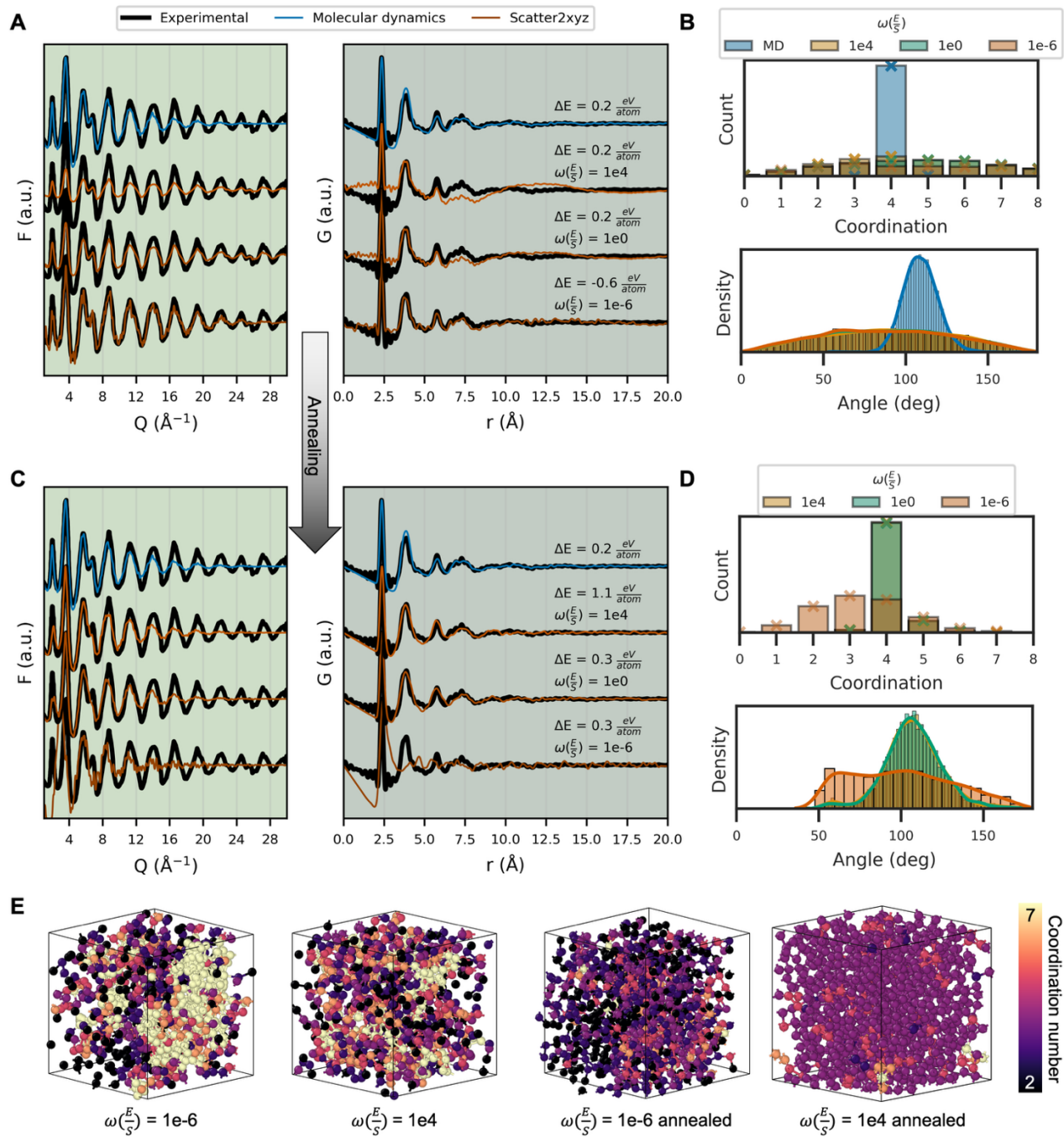

**Figure 5 | Structural interpretation of scattering data for amorphous materials.** (**A**) Comparison of TS and PDF data for an MD-generated a-Si structure (blue) and three Scatter2xyz -derived structures (orange), obtained using varying relative weightings of scattering versus energy, based on experimental scattering patterns from ref.(*43*). (**B**) Coordination number and bond-angle distributions for Scatter2xyz models shown in (A). (**C–D**) Improved structural metrics after a 10-ps annealing step of Scatter2xyz structures, highlighting better local bonding environments at the cost of a slightly reduced scattering data match. E) Representative structures colour-coded by coordination number (within a radial cutoff of 2.85 Å) for different scattering-to-energy weightings

($10^{-6}$ scattering-heavy, $10^4$ energy-heavy) before and after annealing. $\omega(\frac{E}{S})$ is the weighting of the energy versus scattering data.

**Discussion**

We have introduced a unified differentiable optimisation framework for interpreting X-ray or neutron scattering data, with a view to integrate this into autonomous laboratories.

In contrast to ML-based scattering data analysis, which is fundamentally constrained by the scope of training sets, differentiable optimisation is inherently agnostic to chemical space. This generality currently carries a computational cost (up to 168 GPU-hours per refinement with the Adam optimiser(*23*)), yet it scales with the number of GPUs (section E, SI), and the underlying framework is expected to readily transfer to faster optimisation algorithms and hardware accelerators. For example, preliminary tests using a limited-memory Broyden–Fletcher–Goldfarb–Shanno optimiser(*44*) delivered a 661-fold speedup, reducing runtime to <15 GPU-mins (section E, SI). Additional computational efficiencies may arise from surrogate ML models for scattering data calculations(*45, 46*) and from ongoing developments in widely applicable ML interatomic potentials(*6, 30*) and automated workflows(*47*), increasingly bridging the gap between standard theoretical assumptions (e.g., 0 K and vacuum) and realistic experimental conditions.(*7, 48*) Together, these arguments establish a unified differentiable optimisation framework as a general and autonomous tool for scattering data interpretation.

Within autonomous experimentation, the framework can continuously interpret incoming data streams and translate them into structural models. Unlike traditional approaches that return a single best-fit model, this method explicitly maps the full structural solution landscape, capturing inherent degeneracies in scattering data. Such mapping provides a direct diagnostic: convergence to a single, chemically sensible minimum indicates sufficient information, whereas persistent degeneracies highlight the need for additional experimental inputs. As demonstrated in the $P_4$ benchmarks (Figure 2), some scenarios (I, III, IV) yield unique outcomes, while others (II, V) remain ambiguous. In an autonomous laboratory context, these outcomes can automatically trigger acquisition of complementary probes or data modalities, allowing the workflow to adapt dynamically to the information content of the experiment rather than stalling when unique solutions cannot be achieved.

A limitation of the present work is that it has focused on single-phase systems using idealised data, and in experimental practice, the inevitable presence of noise will introduce additional challenges. Furthermore, in practical synthesis, mixtures of phases are likely to dominate. Extending the approach to mixed-phase systems is a natural next step, potentially in combination with unsupervised methods to disentangle contributions from different phases.(*49, 50*) While optional chemical constraints currently require some human intervention, many such constraints can be autonomously derived from experimental data streams; for instance, atomic environment similarity constraints from NMR. Taken together, the ability to explore structural landscapes across diverse systems, while integrating multi-modal scattering data and complementary constraints such as interatomic potentials, positions this framework as a future component for autonomous laboratories.

**Data availability**

All data described in this manuscript will be made available upon publication.

**Code availability**

The Scatter2xyz software will be made available upon publication.

A folder providing the specific code, results and figures used for this manuscript will be made available upon publication.

**Acknowledgements**

The authors thank Dr. Shiwei Liu for valuable discussions about the learning rate adjustments, Dr. Mark Spillman for discussions about the particle swarm optimiser, and Dr. Martin Uhrin for insightful input on gradient-based optimisation.

**Funding**

Novo Nordisk Foundation (grant NNF23OC0081359) (ASA)

UK Research and Innovation [grant number EP/X016188/1] (VLD)

UK Research and Innovation [grant number EP/Z534031/1] (ALG)

European Research Council (788144) (ALG)

**Author contributions**

**Conceptualisation**: ASA, ALG, VLD;

**Methodology**: ASA;

**Investigation**: ASA, JLAG, LAMR;

**Software**: ASA, JLAG, LAMR;

**Formal analysis**: ASA;

**Visualization**: ASA;

**Writing – original draft**: ASA;

**Writing – review & editing**: Everyone;

**Funding acquisition**: ASA, ALG, VLD.

## Competing interests

The authors declare no competing interests.

Supplementary Information

# Autonomous interpretation of atomistic scattering data

Andy S. Anker[1,2]*, John L. A. Gardner[1], Louise A. M. Rosset[1], Andrew L. Goodwin[1], Volker L. Deringer[1]

[1]Department of Chemistry, University of Oxford, Oxford OX1 3QR, United Kingdom

[2]Department of Energy Conversion and Storage, Technical University of Denmark, DK-2800 Kgs. Lyngby, Denmark

*Corresponding author. Email: andy.anker@chem.ox.ac.uk / ansoan@dtu.dk

Table of Contents

**Method: Scatter2xyz**

<u>Scatter2xyz implementation</u>

The framework begins by combining experimental scattering datasets (small-angle scattering (SAS), powder diffraction (PD), or total scattering (TS) with pair distribution function (PDF) analysis) with chemical constraints, subsequently generating multiple candidate atomic structures—termed *walkers*—typically initialised from random atomic coordinates, unit cells, and compositions (Figure 1, initialisation). Each walker is then refined through an iterative two-stage process:

*Gradient-based refinement*

Each walker is optimised independently using a differentiable loss function that combines three components (further details in section *Loss terms and constraints*):

- o   The goodness-of-fit between simulated and experimental scattering data,

- o   Chemical constraints (e.g. minimum distances, similarity metric, or symmetry),

- o   and energy contributions from interatomic potentials.

Learning-rate scheduling (e.g. CosineAnnealingWarmRestarts) cyclically alternates between higher rates, promoting global exploration, and lower rates, enabling precise local refinement (section B, SI).

*Particle swarm update*

The walker with the lowest loss value is used to guide others by shifting their configurations closer to this optimum. Iteration between the local and global phases progressively drives the ensemble of walkers towards improved structural agreement with data and constraints (section B, SI).

<u>Scattering intensity calculations</u>

Scattering intensities were calculated using a fully differentiable implementation of the Debye scattering equation (Eq. 1):*(1, 2)*

Eq. 1
$$I(Q) = \sum_{\nu=1}^{N} \sum_{\mu=1}^{N} b_\nu b_\mu \frac{\sin (Q r_{\nu\mu})}{Q r_{\nu\mu}}$$

Here, $Q$ is the momentum transfer defined by wavelength $\lambda$ and scattering angle $\theta$ (Eq.2):

Eq. 2
$$Q = \frac{4\pi sin(\theta)}{\lambda}$$

Where $N$ is the number of atoms in the structure and $r_{\nu\mu}$ is the distance between atoms $\nu$ and $\mu$. For X-ray radiation, the atomic scattering factor, $b$, depends strongly on $Q$ and is usually denoted as $f(Q)$, but for neutrons $b$ is independent of $Q$ and referred to as the scattering length. For X-rays, the Q-dependency of the atomic scattering factor is approximated using the Cromer-Mann coefficients (Eq. 3):*(3-5)*

Eq. 3
$$b(Q) = \sum_{i=1}^{4} a_i * \exp\left(-b_i * \left(\frac{Q}{4\pi}\right)^2\right) + c$$

To allow continuous optimisation across atomic numbers, pseudo-scattering factors were introduced by linear interpolation between adjacent elements.

Scattering data conversions

Measured intensities can be transformed to the total scattering structure function $S(Q)$, the reduced structure function $F(Q)$, and real-space pair distribution functions $G(r)$ following standard relations (Eqs. 4–6).

Eq. 4
$$S(Q) = \frac{I_{coh}(Q) - \langle b(Q)^2 \rangle + \langle b(Q) \rangle^2}{N \langle b(Q) \rangle^2}$$

Eq. 5
$$F(Q) = Q * (S(Q) - 1)$$

Eq. 6
$$G(r) = \frac{2}{\pi} \int_{Q_{min}}^{Q_{max}} F(Q)\, sin(Q \cdot r)\, dQ$$

These conversions provide access to real-space atomic distance distributions, enabling direct comparison with experimental PDFs. The simulation parameters are given below.

| | $P_4$ | Crystalline Si | Crystalline $CeO_2$ | Arbitrary disordered structures | $C_{60}$ |
|---|---|---|---|---|---|
| $Q_{min, SAXS}$ (Å$^{-1}$) | 0.01 | – | – | 0.01 | 0.01 |
| $Q_{max, SAXS}$ (Å$^{-1}$) | 3.00 | – | – | 3.00 | 3.00 |
| $Q_{step, SAXS}$ (Å$^{-1}$) | 0.01 | – | – | 0.01 | 0.01 |
| $Q_{min}$ (Å$^{-1}$) | 1.0 | 1.5 | 1.5 | 1.0 | 1.0 |
| $Q_{max}$ (Å$^{-1}$) | 30.0 | 18.49 | 18.49 | 30.0 | 30.0 |
| $Q_{step}$ (Å$^{-1}$) | 0.05 | 0.01 | 0.01 | 0.05 | 0.05 |
| $r_{min}$ (Å) | 0.0 | 0.0 | 0.0 | 0.0 | 0.0 |
| $r_{max}$ (Å) | 30.0 | 60.0 | 60.0 | 30.0 | 30.0 |
| $r_{step}$ (Å) | 0.01 | 0.01 | 0.01 | 0.01 | 0.01 |
| Cutoff (Å) | – | 15 | 15 | – | – |

**Table S1 | Simulation parameters of the simulated data that are analysed in the paper.** All isotropic atomic displacement parameters are fixed to 0.3 Å$^2$.

Energy calculations

Energy terms were evaluated using machine learning (ML)-based interatomic potentials, which provide differentiable forces with respect to atomic coordinates. Energies were not considered differentiable with respect to atomic numbers.

For the conceptual $P_4$ example, a MACE potential(*6, 7*) was trained on 4,280 structures of randomly generated $P_{1-4}$ clusters. In this dataset 1–4 P atoms were randomly placed in a 5 Å × 5 Å × 5 Å box (without enforcing minimum interatomic distance constraints, allowing atoms to occupy arbitrarily close positions) or as diatomic pairs with distances of 1.4–2.6 Å in a 10 Å vacuum. The validation and test sets each included 534 structures. Performance metrics were 9.9 meV atom$^{-1}$ (energy) and 115.4 meV Å$^{-1}$ (forces) on training, with validation errors of 11.5 meV atom$^{-1}$ (energy) and 221.9 meV Å$^{-1}$ (forces), and test errors on 17.6 meV atom$^{-1}$ (energy) and 230.6 meV Å⁻¹ (forces). These levels were sufficient to guide optimisation and ensure physically meaningful evaluations. For the energy calculations of in section C, SI, all atoms were converted to P.

For the a-Si example, the quantum-mechanically accurate, ML-driven Moment Tensor Potential(*8*) model of Ref.(*9*) was employed.

<u>Data normalisations</u>

Unless otherwise specified, scattering data were used in absolute counts. SAS data were log-transformed to balance contributions across the Q-range. When normalisation was activated, all patterns were rescaled to unit maximum intensity. For SAS specifically, this included softplus transformation (removing negative values) and logarithmic scaling, before final rescaling to unity.

<u>Loss terms and constraints</u>

The unified loss combines multiple contributions, weighted according to user-defined parameters:

*Scattering loss*: mean-squared error between simulated and measured intensities (Eq. 7).

$$\text{Eq. 7} \qquad \mathcal{L}_{scattering}\left(I_{scattering}^{measured}, I_{scattering}^{target}\right) = (I_{scattering}^{measured} - I_{scattering}^{target})^2$$

*Energy loss*: Huber loss minimising deviations from a target energy (Eq. 8).

$$\text{Eq. 8} \qquad \mathcal{L}_{energy}(E^{target}, E^{structure}) = \frac{1}{2}min\{|\Delta E|, \beta\}^2 + \beta(|\Delta E| - min\{|\Delta E|, \beta\}),$$

Where $\Delta E = E^{structure} - E^{target}$ and $\beta$ is a threshold parameter controlling the transition from quadratic to linear behaviour.

*Similarity constraint loss*: minimisation of deviations in local environments for atoms of the same type (with atomic numbers rounded to the nearest integer). The similarity score is computed based on the deviation of pairwise distances from their average within a given cutoff distance $d_{cutoff}$.

The similarity score $\mathcal{L}_{sim}$ is calculated as follows (Eq. 9):

Eq. 10
$$\mathcal{L}_{sim}(r_{cutoff}, \alpha, D) = \frac{1}{M} \sum_{\alpha} \frac{1}{N_{\alpha}} \sum_{i} \sum_{j} (d_{ij}^{filtered} - \overline{d_{\alpha}^{filtered}})^2$$

Here, $\boldsymbol{D}$ is the matrix of pairwise Euclidean interatomic distances $d_{ij}$ between atoms $i$ and $j$. The summation over $\alpha$ accounts for each unique atomic species in the structure, where $M$ denotes the total number of atomic species present. For a given atomic species $\alpha$, $N_{\alpha}$ represents the number of pairwise distances between atoms that fall within the cutoff distance $d_{cutoff}$. The term $d_{ij}^{filtered}$ corresponds to the pairwise distances that satisfy the condition $d_{ij} < d_{cutoff}$, while $\overline{d_{ij}^{filtered}}$ represents the average pairwise distance for species $\alpha$ within this cutoff. The similarity score $\mathcal{L}_{sim}$ thus quantifies the deviation of pairwise distances from this average, with the aim of minimising these deviations for atoms of the same species.

*Distance constraint loss*: designed to penalise deviations of interatomic distances that fall outside the allowable range, i.e., distances smaller than a user-defined minimum distance $d_{min}$ or greater than a user-defined maximum distance $d_{max}$. We define the loss function with an exponential penalty for both cases. The loss function is expressed as (Eq. 10):

Eq. 10
$$\mathcal{L}_{dist}(d_{min}, d_{max}, \boldsymbol{D}) = \sum_{i,j} \left[ \exp(\max(0, d_{min} - d_{ij})) - 1 \right] + \sum_{i,j} \left[ \exp(\max(0, d_{ij} - d_{max})) - 1 \right]$$

*Symmetry constraint loss*: penalty enforcing cubic cell metrics (Eq. 11)

Eq. 11
$$\mathcal{L}_{cubic}(a, b, c, \alpha, \beta, \gamma) = (a - b)^2 + (b - c)^2 + (c - a)^2 + (\alpha - 90°)^2 + (\beta - 90°)^2 + (\gamma - 90°)^2$$

Where $(a, b, c)$ are the cell-edge lengths and $(\alpha, \beta, \gamma)$ the cell angles.

*Unified loss* as (Eq. 12):

Eq. 12
$$\mathcal{L}_{unified} = \sum \omega_{\text{P}} \cdot \mathcal{L}_{\text{P}}$$

where each term $\mathcal{L}_P$ was weighted by its respective factor $\omega_P$.

R-factor

To evaluate fit quality, the weighted-profile R-factor was computed following crystallographic conventions (Eq. 13):(*10-12*)

Eq. 13
$$R_{wp}^2 = \sqrt{\frac{\sum_{i=1}^n [I_{obs}(Q_i) - I_{calc}(Q)]^2}{\sum_{i=1}^n I_{obs}(Q_i)^2}} \cdot 100\ \%$$

where $I_{obs}(Q)$ is the ground-truth PD pattern in Q space, and $I_{calc}(Q)$ is the PD pattern in the Q-space simulated from the predicted structural model. We assign an equal weight to every point.

**A: Conceptual hierarchy of methods for scattering data analysis**

Reverse Monte Carlo (RMC): a data-driven approach where modelling is iteratively adjusting atomic positions to match experimental scattering data.(*13, 14*)

Empirical potential structure refinement (EPSR) and hybrid RMC: similarly refine structures against scattering data while incorporating interatomic potentials, improving physical plausibility.(*15, 16*)

Molecular dynamics (MD): A physics-based approach where atomic trajectories evolve under interatomic potentials to explore thermodynamically and kinetically accessible structures.(*17*)

Grand-canonical MD: An extension of MD in which atom numbers and/or composition can fluctuate during the simulation.(*18*)

Augmented MD: Multi-modal extension of MD where energies and forces are augmented with experimental data to guide the dynamics.(*19, 20*)

Small-box modelling: refines user-defined structural models against scattering data, allowing extensive chemical knowledge integration.(*21, 22*)

<u>Complex modelling</u>: Extended small-box modelling by combining multiple experimental and computational modalities.(*23*)

<u>ML analysis</u>: In these methods, models are first trained on large datasets of atomic structures paired with simulated scattering patterns. Once trained, the model can map experimental data onto sets of atomic coordinates.(*24, 25*)

<u>Multi-modal ML analysis</u>: Extended ML analysis by combining multiple experimental and computational modalities.(*26*)

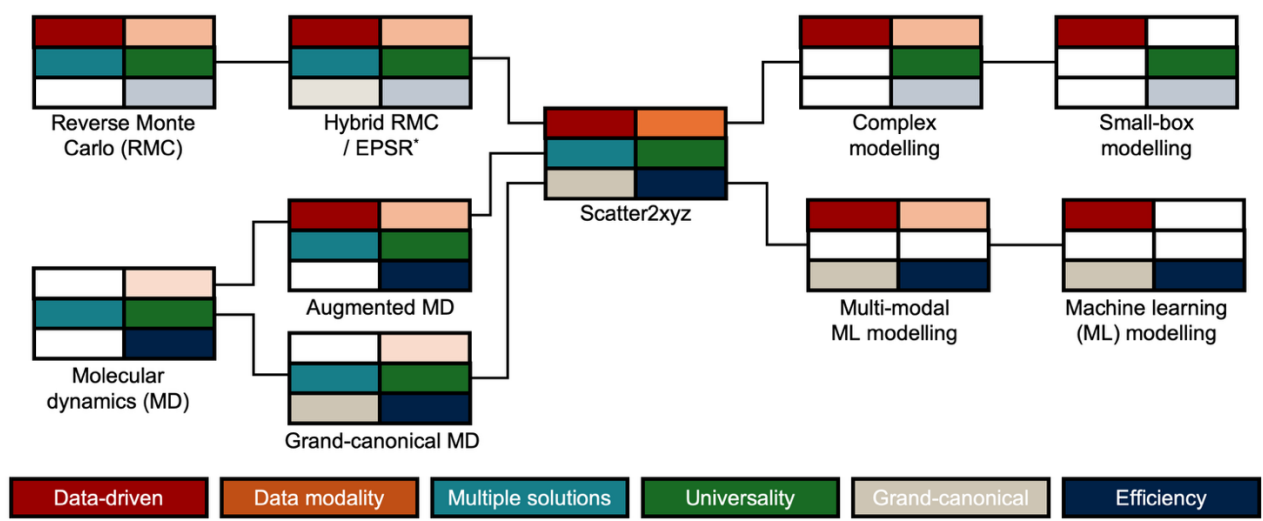

**Figure S1 | A conceptual hierarchy of methods for scattering data analysis.** This schematic qualitatively illustrates whether methods are data-driven (red), capable of integrating multi-modal inputs (orange), able to suggest multiple solutions (teal), universally applicable rather than targeted to specific databases (green), capable of operating under grand-canonical conditions (beige), and their relative optimisation efficiency (blue). Note that this representation is inherently subjective and intended only as an approximate comparison. Certain methods or their specific implementations may exhibit additional capabilities beyond those indicated here.

**B: Optimisation strategies**

In the optimisation process, structural updates occur through backpropagation, with cycles between higher and lower learning rates enabling effective exploration of global and local minima. In an optional secondary optimisation phase, the best-performing walker remains stationary, while the other walkers move their configurations towards it. This combined local and global optimisation scheme effectively navigates complex structural solution landscapes, mitigating the risk of walkers becoming trapped in local minima.

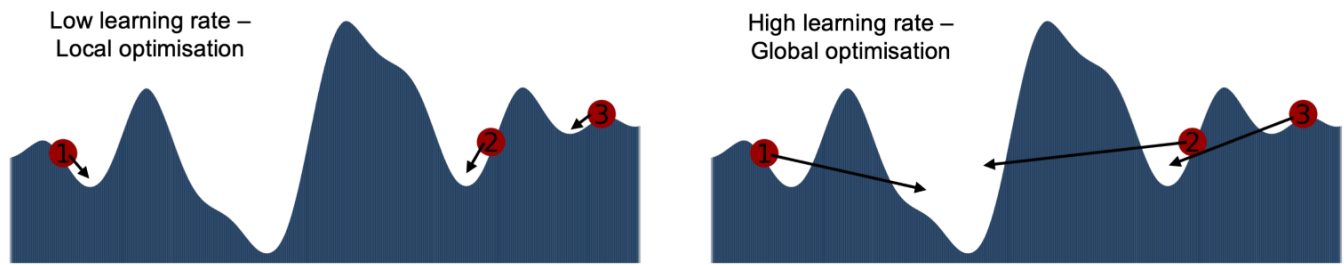

**Figure S1 | Learning rate adjustments.** The figure illustrates that Scatter2xyz is an advanced optimisation scheme in which multiple *walkers* (here, $N = 3$) alternate between high and low learning rates, using the former to escape local minima on the global solution landscape and the latter for local refinement.

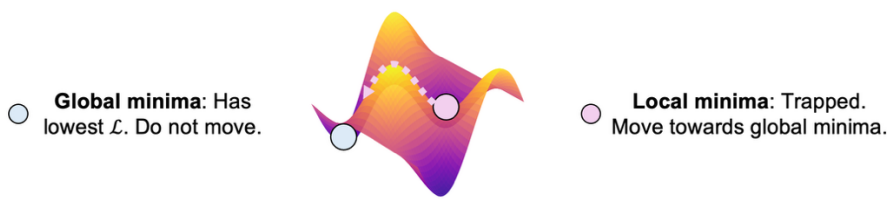

**Figure S2 | Particle swarm optimisation.** Structures located at the global minimum (lowest loss, $\mathcal{L}$) remain stationary, while those trapped in local minima adjust their configurations towards the global minimum. This cooperative update helps walkers escape local minima and converge more efficiently on optimal solutions.

**C: Grand-canonical scattering data interpretation**

Here, optimisation starts from five randomly placed atoms of unspecified species, aiming to identify a four-atom $P_4$ structure by removing one atom entirely. The Scatter2xyz implementation evaluates element identities based on their scattering power, which for X-ray scattering correlates with electron count. To remove an atom, the optimisation progressively lowers its assigned electron count towards zero, represented here as atomic number of zero ($Z = 0$). In the identified global minimum structure, one atom, assigned boron (B, $Z=5$), is progressively being reduced in electron count, indicating ongoing removal. Simultaneously, other atoms are assigned silicon (Si, $Z=14$) and sodium (Na, $Z=11$) closer in electron count to phosphorus (P, $Z=15$). Although this result does not yet reproduce the exact $P_4$ structure, the scattering pattern is well matched, and further optimisation would likely complete the atom removal process.

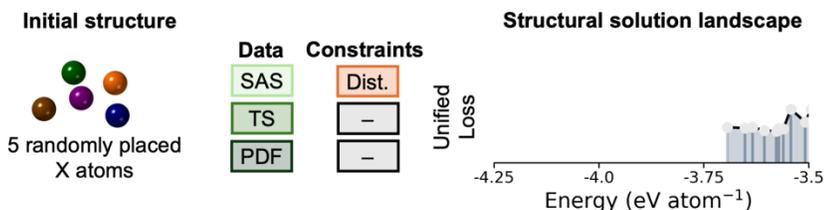

**Figure S3 | Grand-canonical scattering data interpretation.** Starting from five randomly placed atoms of random species, an atom is removed to improve the match to scattering data, exemplifying its capability to optimise composition, atomic positions, and atom count simultaneously.

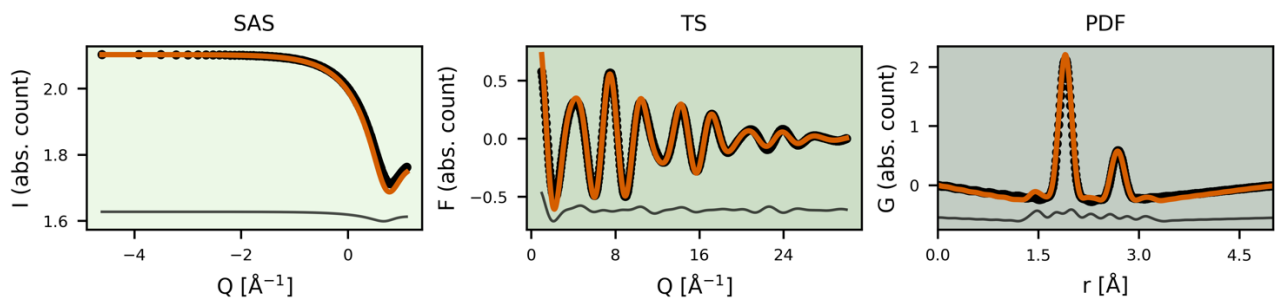

**Figure S4 | Scattering data for the partially converged grand-canonical analysis.** Panels show the SAS, TS, and PDF data (black) overlaid with the corresponding simulated scattering from the optimised structure (orange). Grey lines represent the difference between the two. In this example, the optimisation began with five randomly placed atoms of unspecified species and removed one atom to better match the target scattering pattern, converging on a $BP_2SiNa$ configuration that remains incomplete but closely resembles the ground truth structure. Additional optimisation would likely complete the refinement.

### D: Optimisation of $CeO_2$ structure to the neutron PD data

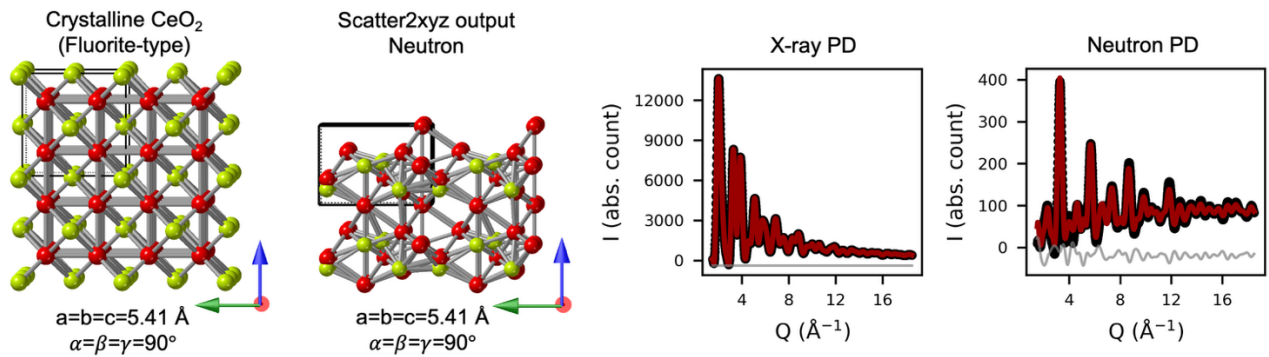

**Figure S5 | Optimisation of $CeO_2$ structure to the neutron PD data.** Structural comparison of the $CeO_2$ (fluorite-type) structure, the optimised structure and corresponding fit to the neutron PD data.

**E: Benchmarking a general approach to scattering-based structural characterisation**

Randomly generated disordered configurations of 4–256 atoms were optimised against simulated scattering data. Atoms were placed at least 2 Å apart within boxes of 3–17 Å, and optimisation was considered successful upon reaching a loss threshold of $5\cdot10^{-5}$. The required computational effort scales strongly with system size: from about one minute for four-atom systems, to hours for tens of atoms, and days for hundreds. Scaling with GPU number, however, allows acceleration through parallelisation. Moreover, benchmarks of a limited-memory Broyden–Fletcher–Goldfarb–Shanno (LBFGS) optimiser[27, 28] indicate potential speedups of more than 600-fold compared to Adam.[29] LBFGS achieves this efficiency by approximating second-order curvature information, enabling larger and more targeted optimisation steps than first-order methods. This benchmark shows that, given sufficient computational resources, scattering data for arbitrary structures can be interpreted without prior training. While a loss below $5\cdot10^{-5}$ does not strictly guarantee uniqueness this can be solved by inclusion of additional experimental modalities and higher data quality.

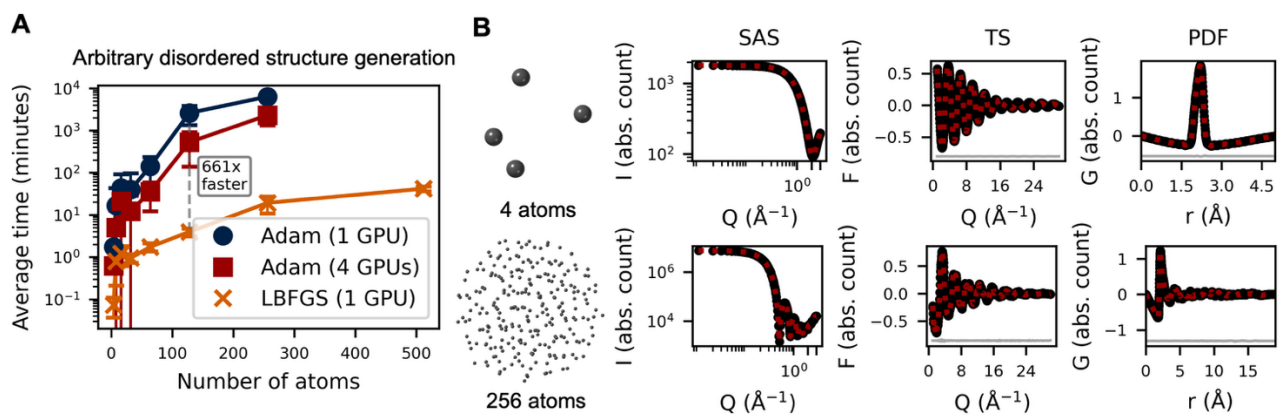

**Figure S6 | Benchmarking a general approach to scattering-based structural characterisation.** A) Computational time required to interpret scattering data from randomly generated atomic structures (defined by achieving a loss below $5\cdot10^{-5}$). Data points represent the average time of five independent runs, with error bars indicating one standard deviation. For the 256 atom, 1 GPU, Adam optimiser run, only three runs out of five finished before the allocated 168 GPU-hours. B) Representative examples of optimised models for disordered systems, showing the agreement between experimental (black) and simulated (red) SAS, TS, and PDF data.

**Supplementary references**